\documentclass[prd, twocolumn,nofootinbib,useAMS]{revtex4}
\usepackage{graphicx,epsfig}

\newcommand{\LCDM}{$\Lambda$CDM\ }

\begin{document}

\title{Dark Energy, A Cosmological Constant, and Type Ia Supernovae}

\author{Lawrence M. Krauss and Katherine Jones-Smith}
\affiliation{CERCA, Case Western Reserve University, Cleveland OH 44106-7079}

\author{Dragan Huterer}
\affiliation{Kavli Institute for Cosmological Physics and Department of Astronomy 
and Astrophysics, University of Chicago, Chicago, IL 60637}

\begin{abstract}
We focus on uncertainties in supernova measurements, in particular of
individual magnitudes and redshifts, to review to what extent supernovae
measurements of the expansion history of the universe are likely to allow us to
constrain a possibly redshift-dependent equation of state of dark energy,
$w(z)$. focus in particular on the central question of how well one might rule
out the possibility of a cosmological constant $w=-1$.  We argue that it is
unlikely that we will be able to significantly reduce the uncertainty in the
determination of $w$ beyond its present bounds, without significant
improvements in our ability to measure the cosmic distance scale as a function
of redshift.  Thus, unless the dark energy significantly deviates from 
$w(z)=-1$ at some redshift, very stringent control of the statistical and systematic
errors will be necessary to have a realistic hope of empirically distinguishing exotic dark energy from
a cosmological constant.

\end{abstract}

\maketitle

\section{Introduction}
Eight years ago two teams observing distant Type Ia supernovae (SNe Ia) announced
evidence that the expansion of the universe is speeding up \cite{Riess_98,
Perlmutter_99}. The distant supernovae appear dimmer than they would be in a
matter-only universe.  If this is a true distance effect, it implies that about
70\% of the energy density of the universe reside in a smooth component with
negative pressure, leaving only about 25\% is in dark matter and 5\% in
baryonic matter.  While compelling evidence for precisely this combination was
pointed out several years earlier \cite{Krauss_Turner}, based on measurements
of the clustering of galaxies, age of the universe, measurements of the baryon
and dark matter densities, and the Hubble constant, SNe Ia were the ``shot
heard around the world'' as they provided explicit evidence for the largest
contributor to the energy density of the universe. This constituent became
known as dark energy.

Since that time the constraints have improved
significantly: combination of cosmic microwave background experiments
\cite{WMAP}, large-scale structure surveys \cite{SDSS,BAO} and, particularly,
new SNe Ia observations \cite{Riess_04,Knop,Astier,Riess_06,ESSENCE}, now
constrain the dark energy equation of state, $w\equiv p/\rho$ to be within
about $-1\pm 0.1$.

Unfortunately, our understanding of dark energy is as murky as ever.  The
simplest model for dark energy is provided by the cosmological constant term in
Einstein's equations, and is still an excellent fit to observations as it
predicts $w=1$ identically and at all times.  However, even if observations
were to pin down the value of $w$ to be $-1$, this gives us essentially no
insight into the possible source of dark energy.  While a cosmological constant
may be the best bet, it is simple to imagine other sources of dark energy,
including the energy density associated with a false vacuum metastable scalar
field, that would produce a similar value.  The only way to get any new
theoretical handle on dark energy is to be able to unambigiously determine a
deviation from $-1$, if such a deviation indeed exists.

Various proposals for dynamical dark energy have been put forth which might
produce such deviations.  Most notably, a scalar field rolling down its
effective potential can provide the necessary energy density and acceleration
of the universe \cite{Wetterich,Freese,Ratra,
Coble,Ferreira_Joyce,Caldwell}. Generically many of these possibilities are already
ruled out by the data.  Among those that are not, none is particularly
compelling as they also typically do not naturally address the problem why the
observed dark energy density ($\sim 10^{-120} M_{\rm PL}^4$) is so small, nor
why dark energy starts to dominate the expansion of the universe only at recent
times --- redshift $z \lesssim 1$.  Nevertheless, before we can address such
puzzles we need to know empirically if the dark energy is measurably
distinguishable from a cosmological constant.

Here we will be concerned with two specific observational factors, as well as
one overriding theoretical constraint.  Observationally we need to determine
both the magnitude and redshift of individual objects in order to map the
universe's expansion history.  Theoretically we have to account for the fact
that dark energy has, a priori, no predetermined time dependence, and thus our
analyses must allow for arbitrary time variations.

Essentially all of the consequences of dark
energy follow from its effect on the expansion rate:

\begin{eqnarray}
\label{eq:H^2}
H^2 & = &  {8\pi G \over 3} (\rho_M + \rho_{\rm DE}) \\[0.1cm]
H^2(z)/H_0^2  &  = & \Omega_M(1+z)^3 +  \nonumber\\
    && \Omega_{\rm DE} \exp \left[ 3\int_0^z\,[1+w(z')]d\ln (1+z') \right] \nonumber
\end{eqnarray}

\noindent where $\Omega_M$ and $\Omega_{\rm DE}$ are the dark matter and dark
energy density relative to critical, respectively, and we have ignored the
relativistic components.  Type Ia supernovae effectively measure the luminosity
distance

\begin{equation}
d_L(z)=(1+z) r(z)=\int_0^z {dz'\over H(z')},
\end{equation}

\noindent where $r(z)$ is the comoving distance and we have assumed a flat universe.
Since observations of supernovae allow us, at least in principle, to map the
expansion history of the universe, one can use this data to constrain the
nature of dark energy.  

This paper is organized as follows. In Sec.~\ref{sec:expansion}, we review a variety of 
parametrizations of the expansion history of the universe, and therefore ways to
measure the properties of dark energy. In Sec.~\ref{sec:testlambda} we study the extent
to which an increased statistical error in supernova distances affects determination
of the equation of state of dark energy and tests of its consistency with the vacuum
energy value of $-1$. We conclude in Sec.~\ref{sec:conclusions}.

\section{Mapping the expansion history}\label{sec:expansion}

\subsection{Key questions and parametrizations}

At the present time, it has become clear that there are two major goals that
upcoming dark energy probes should address:

\begin{enumerate}
\item Is dark energy consistent with the vacuum energy scenario -- that is, is
$w(z)=-1$?

\item Is the equation of state $w$ constant in time (or redshift)?
\end{enumerate}

Violation of either of these two hypotheses would be a truly momentous discovery:
the former would rule out a pure cosmological constant, while
the latter would provide further evidence for nature of of dark energy via its dynamics.

With this in mind, the most obvious approach is to parametrize the equation of state
of dark energy as a constant piece plus a redshift-varying one \cite{Cooray_Huterer,
Linder_w0wa, Bassett}

\begin{eqnarray}
w(z)  &=& w_0 + w' z       \label{eq:wz_CooHut} \\[0.1cm]
w(z)  &=& w_0 + w_a z/(1+z) \label{eq:wz_Linder} \\[0.1cm]
w(z)  &=& w_0 + {w_f-w_0\over 1+ \exp[(z-z_t)/\Delta ]}  \label{eq:wz_CorCop} 
\end{eqnarray}

\noindent where the first equation assumes linear evolution with redshift, the
second is linear with scale factor, and the third allows for the transition
between two asymptotic constant values of the equation of state, with the
transition at redshift $z_t$ with the characteristic width in redshift
$\Delta$. Equation \ref{eq:wz_Linder}, in particular, has become commonly used
to plan probes of dark energy as it retains the minimally required two
parameters and does not diverge at high $z$, while still allowing for the
low-redshift dynamics (e.g.\ variation in the value of $w(z)$). 

Many other simple parametrizations of the equation of state have been suggested
(e.g.\ \cite{Corasaniti_Copeland, Linder_howmany}); similar proposals have been
extended to the Hubble parameter (e.g.\ \cite{Saini_Hz}). Finally, specific
combinations of the expansion history parameters, such as the ``statefinder''
\cite{statefinder}, have been proposed as good discriminators between
phenomenological dark energy descriptions. It is well worth emphasizing that
all of the above parametrizations are ad hoc, and may lead to biases as the
true DE model may not follow the form imposed by these functions. Their
advantage, however, is in simplicity and the fact that two or three additional
parameters in the dark energy sector will be, in the near future, measured to a good
accuracy, at least when the data from the various cosmological probes is
combined.



\subsection{Direct reconstruction}

Going in the opposite direction, the most general way to probe the background
evolution of dark energy has been proposed in \cite{reconstr,Nakamura_Chiba,Starobinsky}:
the equation of the distance vs redshift can be inverted to obtain $w(z)$ as a
function of the first and second derivatives of the (SNe Ia-inferred, for example)
comoving distance

\begin{equation}
1+w(z) = {1+z\over 3}\, {3H_0^2\Omega_M(1+z)^2 + 2(d^2r/dz^2)/(dr/dz)^3\over
        H_0^2\Omega_M(1+z)^3-(dr/dz)^{-2}}
\label{eq:wz_reconstr}
\end{equation}

\noindent Similarly, assuming that dark energy is due to a single rolling
scalar field, one can reconstruct the potential of the field exactly

\begin{eqnarray}
V[\phi (z)] 
&=& {1\over 8\pi G}\left[ {3\over (dr/dz)^2}
        +(1+z) {d^2r/dz^2\over (dr/dz)^3}\right]  \nonumber\\[0.1cm]
&-& {3\Omega_MH_0^2  (1+z)^3 \over 16\pi G} 
\label{eq:Vphi_reconstr}
\end{eqnarray}

\noindent where the upper (lower) sign applies if $\dot\phi >0$ ($<0$) The sign
is arbitrary, as it can be changed by the field redefinition, $\phi
\leftrightarrow -\phi$. 

\begin{figure*}[!t]
\includegraphics[height=2.3in, width= 2.2in, angle=0]{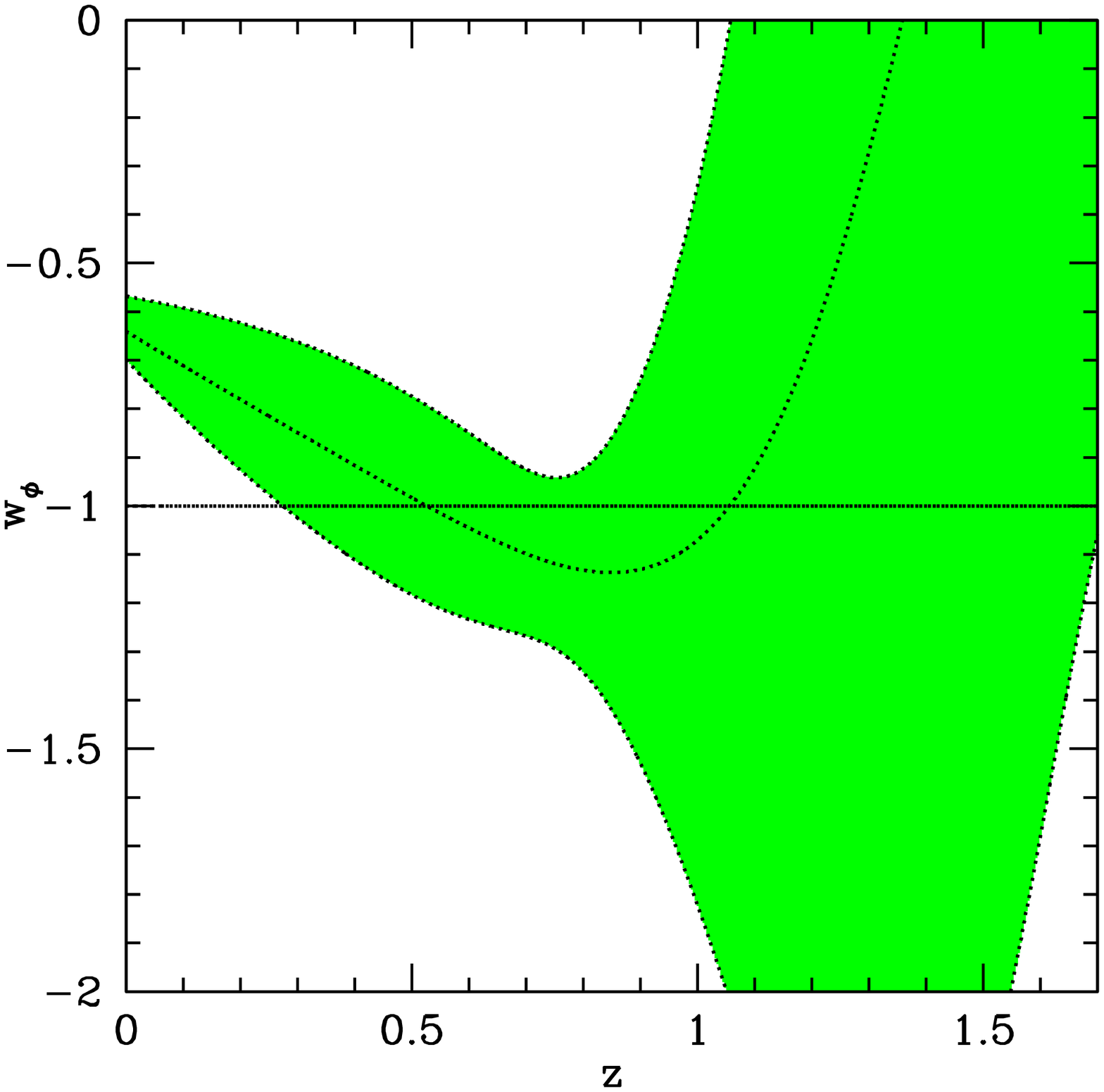}
\includegraphics[height=2.3in, width= 2.2in, angle=0]{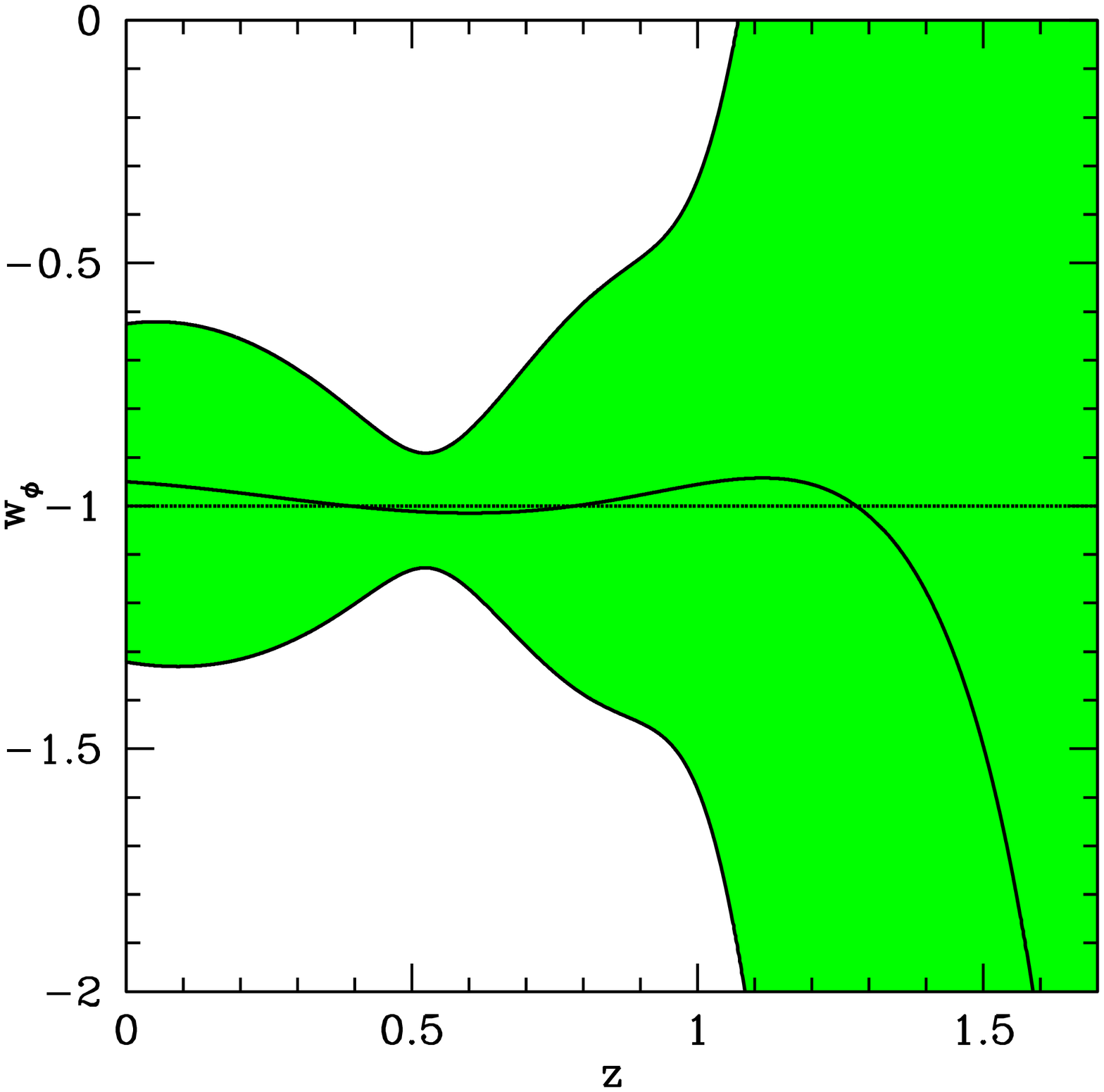}
\includegraphics[height=2.3in, width= 2.2in, angle=0]{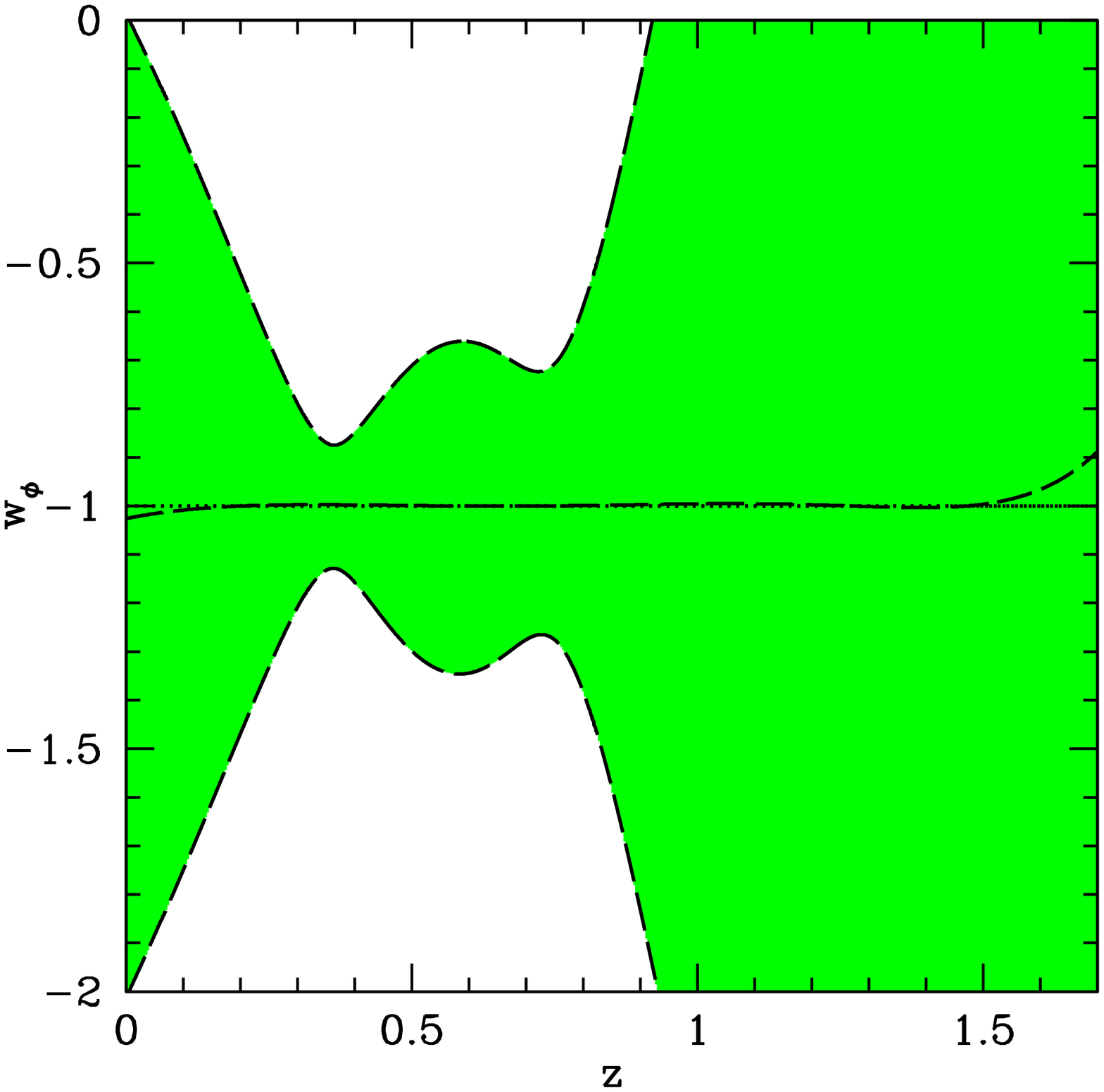}
\caption{Example of the direct reconstruction, simulated for future SNa Ia data
and assuming the true equation of state is $w(z)=-1$. The left, middle and
right panels show results when the luminosity distance data are fit with a
third, fourth and fifth order polynomial in redshift, respectively.  Note that,
depending on which order polynomial is used to fit the data, significant bias,
or statistical error, or both are introduced. Adopted from Ref.\
\cite{Weller_Albrecht}. }
\label{fig:weller_rec}
\end{figure*}

Direct reconstruction is the most general method for inferring the dark energy
history, and it is the only approach that is truly model-independent (despite
some claims in literature to the contrary). However, direct reconstruction
comes at a steep price --- it calls for taking the {\it second} derivative of
the noisy data.  In order to take the second derivative, one essentially must
fit the luminosity distance (or SNa apparent magnitude) data with a smooth
function --- a polynomial, Pad\'{e} approximant, spline with tension
etc. Unfortunately, the parametric nature of the fitting process introduces
systematic biases.  After valiant attempts to do this using a variety of
methods for smoothing or fitting the data (e.g.\
\cite{Huterer_Turner,Weller_Albrecht,Saini_Hz, Zhao}), various authors found
that direct reconstruction is simply too challenging and not robust even with
SNe Ia data of excellent quality. Figure \ref{fig:weller_rec} shows example of
the direct reconstruction, simulated for future SNa Ia data and assuming the
true equation of state is $w(z)=-1$ (adopted from Ref.\
\cite{Weller_Albrecht}). Note that, depending on which order polynomial is
used to fit the data, significant bias, or statistical error, or both are
introduced.  For an excellent review of the dark energy reconstruction and
related issues, see \cite{sahni_review}.

\subsection{Principal components}

The next most general method, introduced in \cite{Huterer_Starkman} (see also
\cite{Crittenden,Simpson_Bridle}), is to compute the principal components of
the quantity that we are measuring --- the equation of state or energy density of
dark energy. Principal components are the redshift weights (or window
functions) of the function in question, and are uncorrelated by construction.
In this scheme, one simply lets data decide which weights of the function are
measured best, and which ones are measured most poorly.  The principal
components form a natural basis that parameterizes the measurements of any
particular survey.

To compute the principal components, let us parametrize $w(z)$ (same arguments
follow for $\rho(z)$ or $H(z)$) in terms of piecewise constant values $w_i$
($i=1, \ldots, N$), each defined in the redshift range $z_i=[(i-1)\Delta z,
i\Delta z]$ where $\Delta z=z_{\rm max}/N$.  In the limit of large $N$ this
recovers the shape of an arbitrary dark energy history (in practice, $N\gtrsim
20$ is sufficient). We then proceed to compute the covariance matrix for the
parameters $w_i$, plus any other cosmological parameters such as $\Omega_M$,
then marginalize over the latter.  We then have the covariance matrix, $C$, for
the $w_i$.

It is then a simple matter to find a basis in which the parameters are
uncorrelated; this is achieved by simply diagonalizing the inverse covariance
matrix (which is in practice computed directly and here approximated with
$F$). Therefore
\begin{equation}
F\equiv C^{-1}= W^T \Lambda W
\end{equation}

\noindent where the matrix $\Lambda$ is diagonal and rows of 
the decorrelation matrix $W$ are the eigenvectors $e_i(z)$, which
define a basis in which our parameters are
uncorrelated~\cite{hamilton}. The original function can be
expressed as 

\begin{equation}
w(z) = \sum_{i=1}^N \alpha_i\, e_i(z)
\label{eq:w_expand}
\end{equation}

\noindent  where $e_i$ are the ``principal components''. Using the orthonormality
condition, the coefficients $\alpha_i$ can be computed as
\begin{equation}
\alpha_i = \sum_{a=1}^N w(z_a)\, e_i(z_a).
\label{eq:coeff}
\end{equation}

Diagonal elements of the matrix $\Lambda$, $\lambda_i$, are the eigenvalues which
determine how well the parameters (in the new basis) can be measured;
$\sigma(\alpha_i)=\lambda_i^{-1/2}$. We have ordered the $\alpha$'s so that
$\sigma(\alpha_1)\leq\sigma(\alpha_2)\leq\ldots\leq\sigma(\alpha_N)$.

\begin{figure}[!t]
\includegraphics[height=3.5in, width= 2.4in, angle=-90]{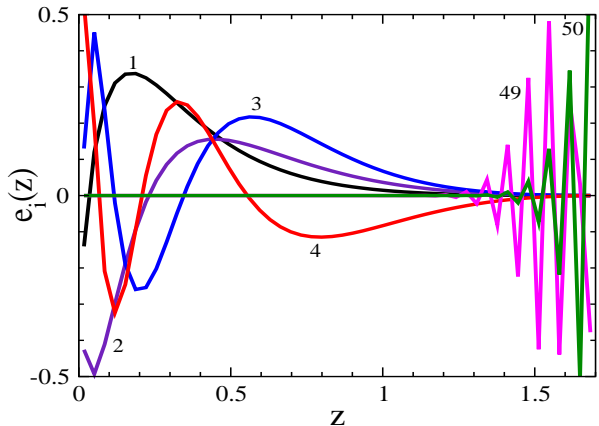}
\caption{The four best-determined and two worst-determined principal components
of $w(z)$ for a future SNe Ia survey such as SNAP \cite{SNAP}. Adopted from
Ref.~\cite{Huterer_Starkman}.  }
\label{fig:PC}
\end{figure}

Principal components have one distinct advantage over fixed
parametrizations. They allow the {\it data} to determine which weight of the
cosmological function $w(z)$ (or $\rho_{\rm DE}(z)$, or $H(z)$) is best
determined. In fact, the PCs depend on the cosmological probe, 
on the specifications of the probe (redshift and sky coverage etc), 
and (more weakly) on the true cosmological model. These dependencies
are features and not bugs, and they make the principal components  a 
useful tool in survey design. For example, one can design a survey 
that is most sensitive to the dark energy equation of state at some specific
redshift, or study how many independent parameters are measured by any given
combination of cosmological probes (e.g.\ \cite{Linder_howmany}).

\subsection{Uncorrelated estimates of DE evolution}

A useful extension of the principal component formalism is to compute the band
powers in redshift of the dark energy function, $w(z)$ (or $\rho_{\rm DE}(z)$ or
$H(z)$). These band powers can be made 100\% uncorrelated by construction, and
the price to pay is a small leakage in the sensitivity of each band power
outside of its redshift range. For details on how this is implemented, see
\cite{Huterer_Cooray}.

\begin{figure}[!t]
\includegraphics[height=3.5in, width= 2.5in, angle=-90]{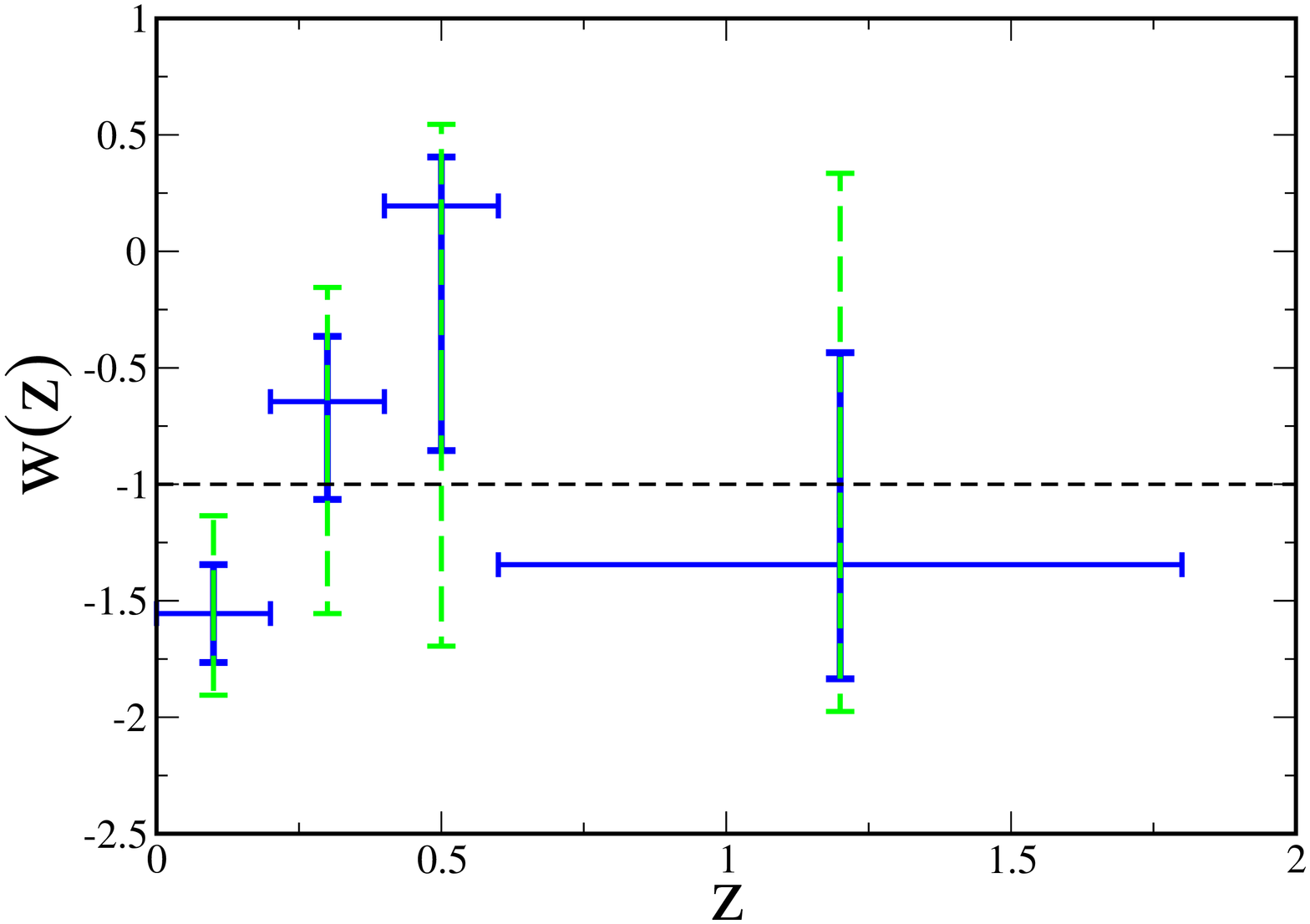}\\
\includegraphics[height=2.1in, width= 3.1in, angle=0]{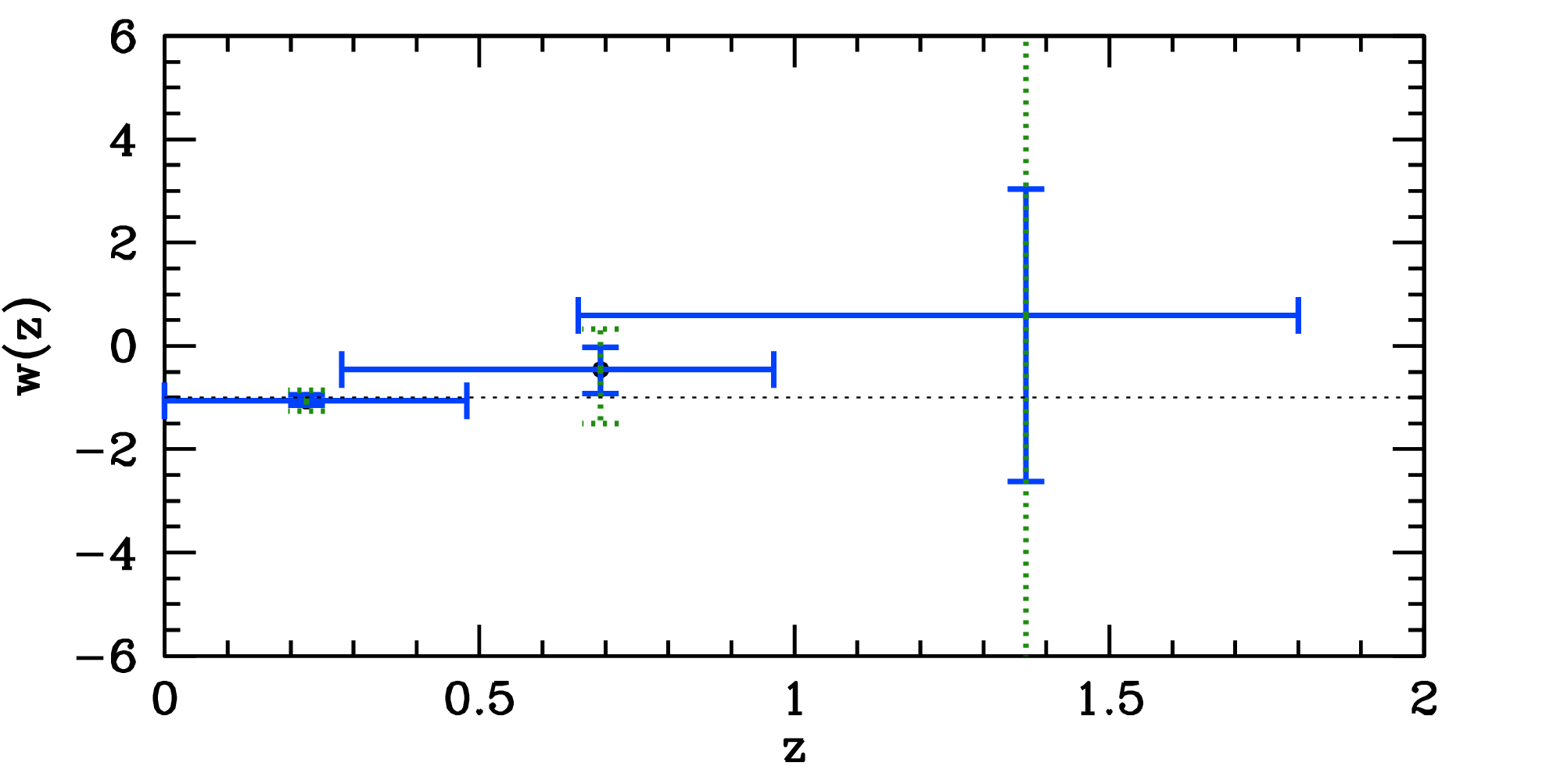}
\caption{Top panel: constraints on the four band powers of $w(z)$, adopted from
Ref.~\cite{Huterer_Cooray}, assuming the Riess04 SNa data \cite{Riess_04} and a
prior on $\Omega_M$. Bottom panel: the same method applied to constrain three
band powers, using the newest data SNa Ia data from the Hubble Space Telescope
(adopted from Ref.~\cite{Riess_06}) in combination with the baryon acoustic
oscillation measurements \cite{BAO}.  }
\label{fig:uncorr}
\end{figure}

The top panel of Figure \ref{fig:uncorr} shows the constraints on the four band
powers of $w(z)$, adopted from Ref.~\cite{Huterer_Cooray}, assuming the Riess04
SNa data \cite{Riess_04} and a prior on $\Omega_M$. The bottom panel of the
Figure shows the same method to constrain three band powers, using the newest
data SNa Ia data from the Hubble Space Telescope, adopted from
Ref.~\cite{Riess_06}, in combination with the baryon acoustic oscillation
measurements \cite{BAO}.. The slight preference for $w(z)$ increasing with
redshift is seen in both panels; however, significant systematics still affect
the data, in particular because of the heterogeneity of the SNa Ia datasets
used. Therefore, it is too early to claim any evidence for the departures from
\LCDM, and it remains to be seen how the results change once we have a
systematically more homogeneous and statistically more powerful data set (for
the requirements on the SNa Ia systematics, see \cite{Kim_Linder}).

It interesting to see in Fig.~\ref{fig:uncorr} how {\it good} the constraints
are, given that this is current data and that interesting constraints are
obtained on 3-4 band powers. A superior data set with an excellent control of
the systematics, such as that expected from a dedicated space telescope such as
SuperNova/Acceleration Probe (SNAP; \cite{SNAP}) will significantly improve the
constraints, and also allow a finer resolution (i.e.\ more band power
parameters) in redshift.

Nevertheless, it is clear from the existing set that either significant
redshift evolution in $w$ or significant improvement in the data will be
required before either redshift evolution or deviation from $-1$ could be
unambiguously inferred using this technique.  We now turn to considering this
question in more detail

\section{Lambda or not?}\label{sec:testlambda}

As stressed earlier, one of the most important outstanding questions in cosmology is
whether dark energy is consistent with a cosmological constant. This
hypothesis can be tested in a variety of ways. The simplest approach is to
compute a simple likelihood comparison between the data and the vacuum energy
model $w(z)=-1$.  A much more sophisticated (but admittedly less robust)
approach would be to perform some type of reconstruction of the energy density
$\rho_{\rm DE}(z)$ and check whether or not it is consistent with a constant value
(or similarly, to reconstruct the equation of state $w(z)$ and check whether it
is consistent with $-1$).

Here we consider the $\Lambda$ hypothesis test in terms of principal components
(PCs).  We consider models in the $w_0$-$w_a$ plane (see Eq.~(\ref{eq:wz_Linder}))
and as we outline below, we use the best measured PCs to determine if we can
statistically distinguish distinguish the model from \LCDM. In this section, we
assume future SNa Ia data with 3000 SNe distributed uniformly in $0.1<z<1.7$;
this corresponds roughly (but not exactly) to what is expected from the SNAP
space telescope \cite{SNAP}.

For a fixed dark energy model described by some values of the cosmological
parameters $w_0$ and $w_a$, let the principal components take values
$\alpha_i$, with associated errors $\sigma(\alpha_i)$ (note, we marginalize the
results over the values of $\Omega_M$, thus enlarging the errors in the PCs; in
this way we can talk about models in the $w_0$-$w_a$ plane without further recourse to
$\Omega_M$).  Further, from Eq.~(\ref{eq:coeff}) it follows that the principal
component coefficients for the $w(z)= -1$ model are

\begin{equation}
\bar{\alpha_i} = (-1)\,\sum_{a=1}^N e_i(z_a),
\end{equation}

\noindent where $e_i(z)$ is the shape of $i$th PC in redshift. 
We can now perform a simple $\chi^2$ test to determine whether
$w$ is constant:
\begin{equation}
\chi^2=\sum_{i=1}^M {(\alpha_i-\bar{\alpha_i})^2
\over \sigma^2(\alpha_i)},
\end{equation}

\noindent where we have chosen to keep only the first $M$ PCs, since the
best-determined modes will contribute the most to the sum (note that the test
is valid regardless of the value of $M$; in practice, our M is typically
3-5). Given that we have $M$ degrees of freedom, each model $(w_0, w_a)$ will
lead to $\chi^2$ which may or may be inconsistent with the \LCDM model at some fixed
confidence level.

In order to determine the ability of future SN surveys to make this
distinction, we focus here on the sensitivity of the results to the measured
magnitude uncertainty.  We focus on this factor for two reason.  First, we
believe it will be the single most important determinant of the ability of
future surveys to possibly rule out a cosmological constant, and second,
because our examination of past surveys suggests that one should consider the
possibility of redshfit-dependent magnitude measurement errors.  This latter
possibility is not unexpected.  It is systematically harder to determine the
luminosity of ever fainter and more distant objects.

Indeed, it is this possibility that originally motivated
\cite{kraussdavislinton} the current study.  Measuring supernovae at ever
higher redshift provide a useful lever-arm to distinguish between different
equations of state, unless the magnitude uncertainty increases more quickly
with redshift than the redshift-distance relation diverges for differing values
of the dark energy equation of state.  For purposes of this example analysis,
we used 217 SN1a, from the initial large SN surveys
{\cite{Perlmutter_99,Riess_98}, supplemented by several more recent
measurements, fitting the quoted magnitude uncertainty as a function of
redshift to a linear relation.  Although the oft-claimed quoted magnitude
uncertainty per low-redshift supernova is $0.15$ we found that a free fit to
the data tended to prefer a slightly larger value.  Thus, for purposes of
comparison, we considered both redshift independent uncertainties of $0.15$ or
$0.21$ magnitudes, as well as redshift dependent uncertainties $0.15+0.079z$ or
$0.21+0.079z$ mag.

Figure \ref{fig:PCtest_manyplots}, shows the regions in the $w_0-w_a$ that are
ruled out at 68\% C.L.\ (region outside of the innermost closed curve) 95\%
C.L.\ (region outside of the middle curve) and 99.7\% C.L.\ (region outside of
the outermost curve, coinciding with the blue shaded region). In other words,
the blue region corresponds to dark energy scenarios where the \LCDM model can
be ruled out at $>3\sigma$ confidence.

In the four panels of Figure \ref{fig:PCtest_manyplots} we show cases when the
error per SNa is $0.15$ or $0.21$ magnitudes, and when it is increasing with
redshift as $0.15+0.079z$ or $0.21+0.079z$ mag.  Clearly, increasing magnitude
uncertainties with redshift can significantly affect the size of the
indistinguishable parts of model space. Note too that the shape of the region
that cannot be distinguished from \LCDM is of characteristic shape, is
elongated roughly along the $7(w_0+1)+w_a\approx 0$ direction --- this is a
direct consequence of the fact that the physically relevant quantity is
$w(z)=w_0 + w_a z/(1+z)$.

\begin{figure*}[!t]
\epsfig{file=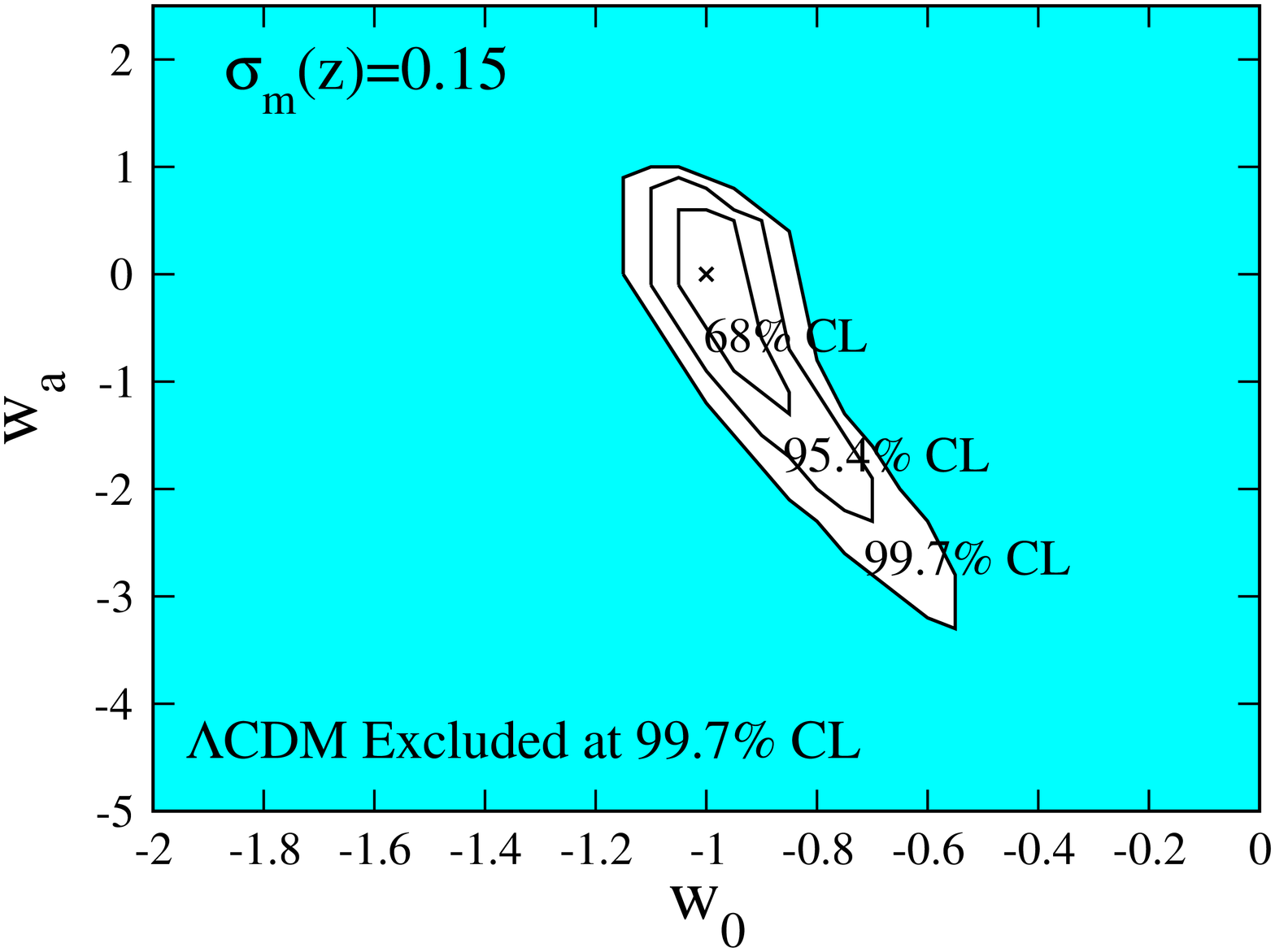,height=3.4in,width=2.8in,angle=-90}
\epsfig{file=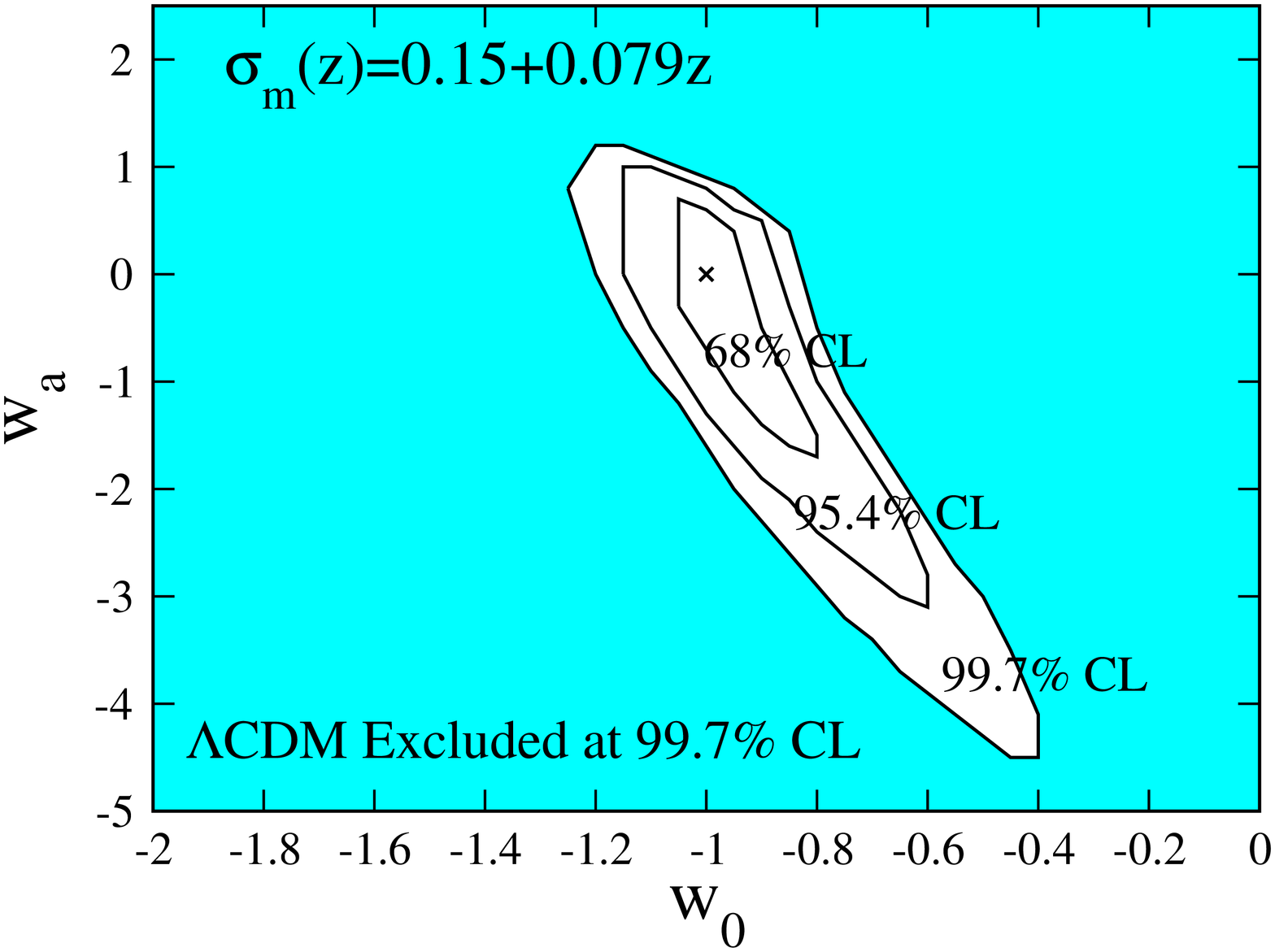,height=3.4in,width=2.8in,angle=-90}\\[-0.2cm]
\epsfig{file=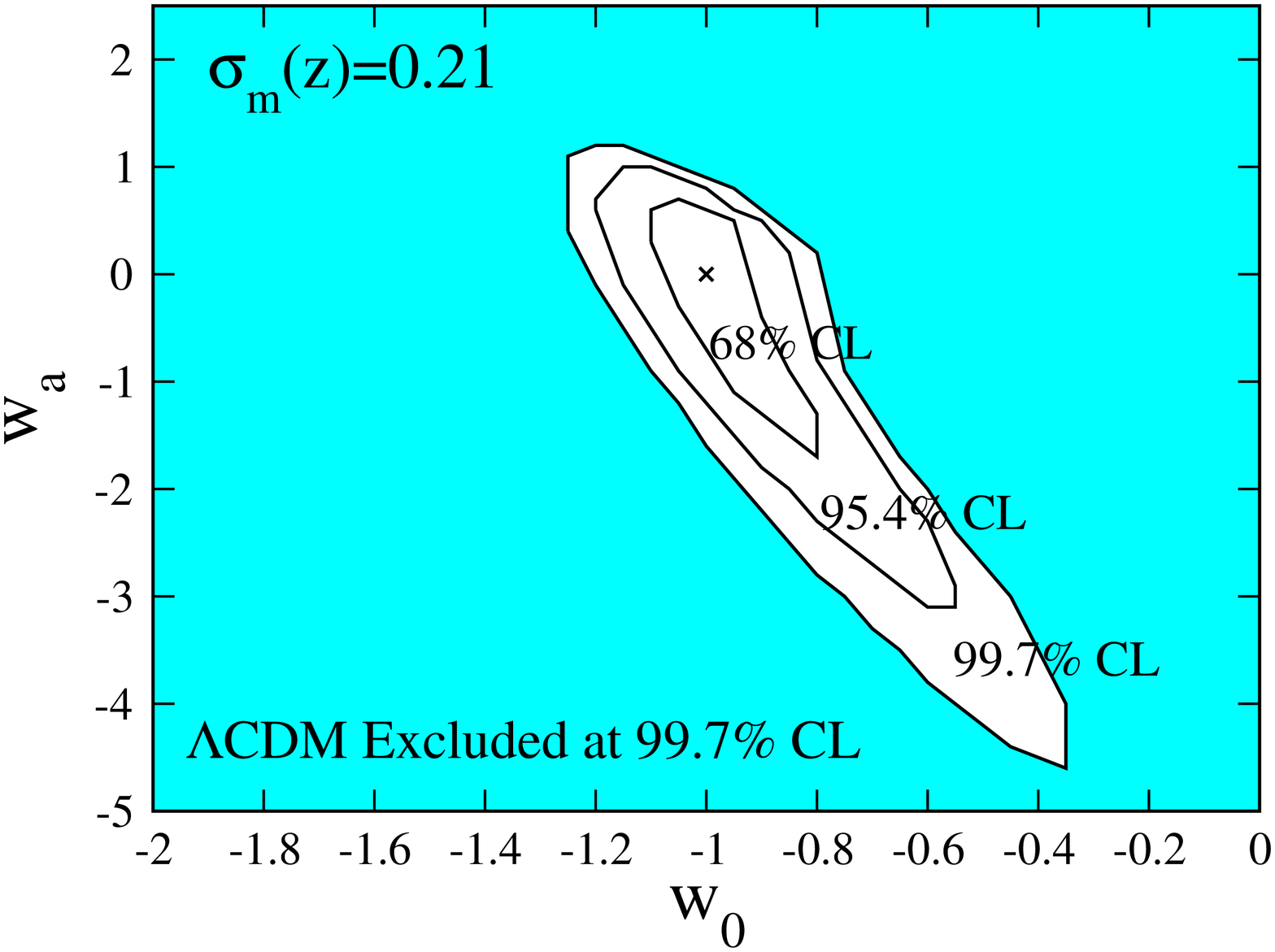,height=3.4in,width=2.8in,angle=-90}
\epsfig{file=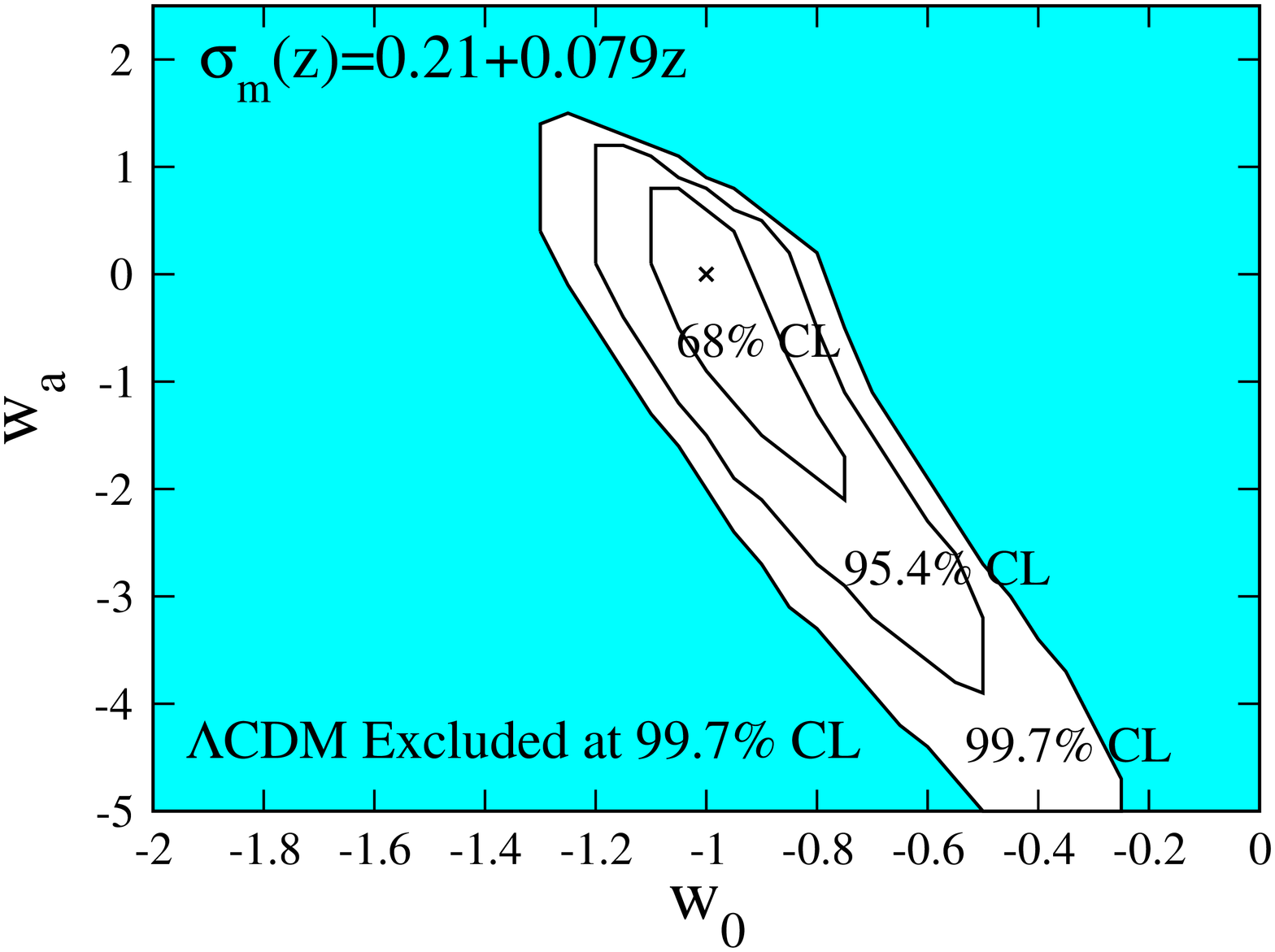,height=3.4in,width=2.8in,angle=-90}
\caption{ Regions in the $w_0-w_a$ that are ruled out at 68\% C.L.\ (region
outside of the innermost closed curve) 95\% C.L.\ (region outside of the middle
curve) and 99.7\% C.L.\ (region outside of the outermost curve, coinciding with
the blue shaded region). In other words, the blue region corresponds to dark
energy scenarios where the \LCDM model can be ruled out at $>3\sigma$
confidence.  This test is performed by ``measuring'' the principal components
from each $(w_0, w_a)$ model and comparing them to those from the \LCDM model.
We show cases when the error per SNa is $0.15$ or $0.21$ magnitudes (top and
bottom left panels), and when it is increasing with redshift as $0.15+0.079z$ or
$0.21+0.079z$ mag (top and bottom right panels).  }
\label{fig:PCtest_manyplots}
\end{figure*}

One other way to get a handle on whether $w \ne - 1 $ is to explore the
most precise constraint on $w$ that might be obtainable {\it at any single
redshift}.  For, if $w \ne - 1 $ at any redshift, this will establish
unambiguously that the dark energy is not a cosmological constant.

To explore the sensitivity of SNe for this purpose
we parametrize dark energy equation of state as in
Eq.~(\ref{eq:wz_Linder}), and study the accuracy in the equation of state $w(z)$
at the best determined, or pivot, point. The pivot value is given by
\cite{Huterer_Turner,DETF}

\begin{equation}
w_{\rm pivot} = w_0 - {{\rm Cov(w_0, w_a)}\over {\rm Cov(w_a, w_a)}}\, w_a
\label{eq:wp_from_w0wa}
\end{equation}

\noindent where ${\rm Cov}(x, y)$ stands for elements of the covariance matrix element,
while the pivot redshift at which $w(z)$ is best determined is given by

\begin{equation}
z_{\rm pivot} = - {{\rm Cov(w_0, w_a)}\over {\rm Cov(w_0, w_a)}+{\rm Cov(w_a, w_a)}}.
\label{eq:zp}
\end{equation}

\noindent Other parameter we are the matter density relative to critical,
$\Omega_M$ and the offset in the SNa Ia Hubble diagram,
$\mathcal{M}$. Throughout we assume a flat universe.

\begin{figure}[!t]
\includegraphics[height=3.6in, width= 2.6in, angle=-90]{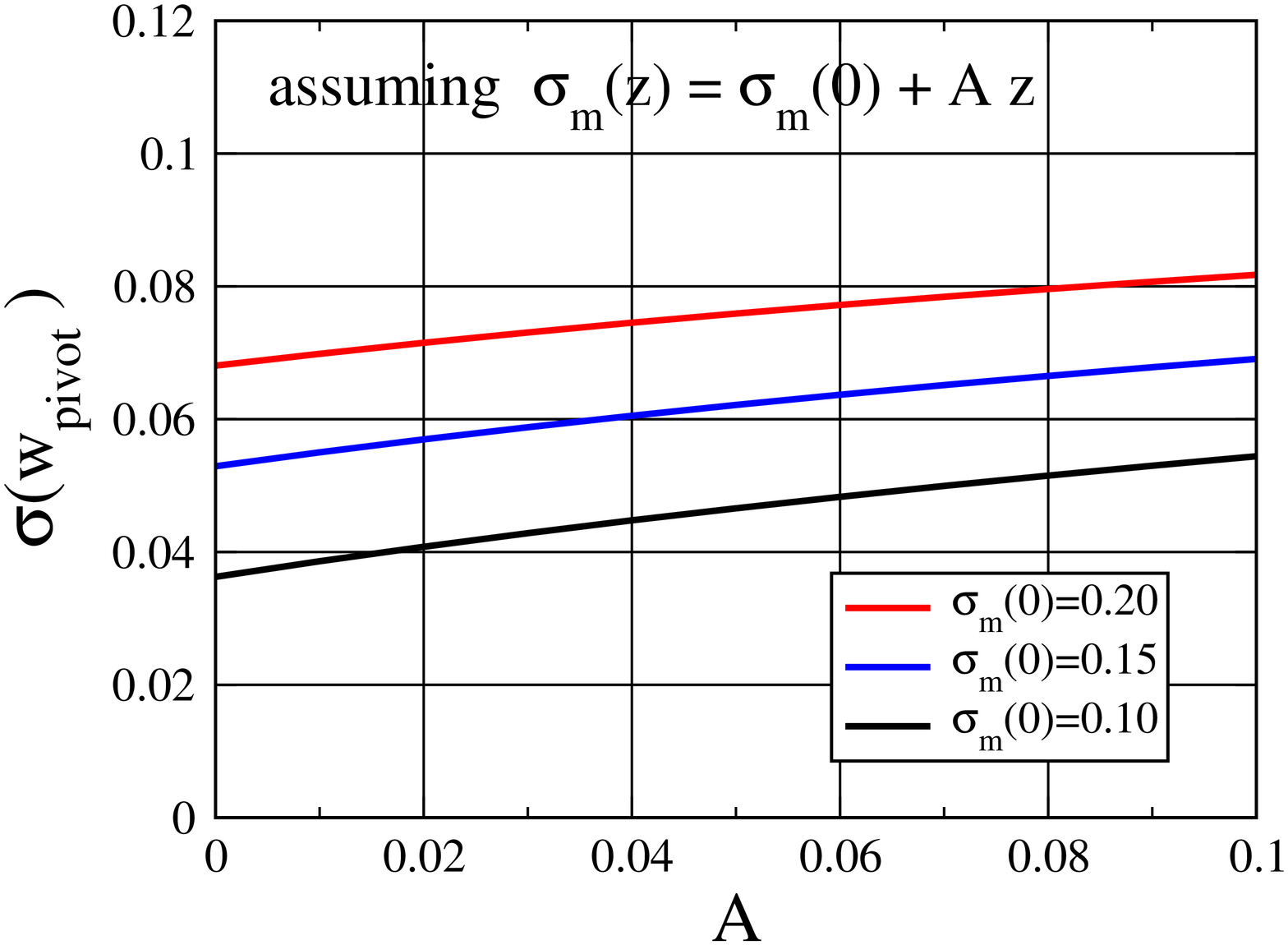}
\caption{Constraints on the pivot value of the equation of state, 
$w_{\rm pivot}$, as a function of the intercept and slope of the magnitude
error of SNe, $\sigma_m(0)$ and $A$. We have assumed a fiducial survey of
2800 SNe. }
\label{fig:sig_wpivot_vs_A}
\end{figure}

Using a Fisher matrix formalism, we estimate errors for a future SNAP-type
survey \cite{SNAP} with 2800 SNe distributed in redshift out to $z=1.7$ as
given by \cite{SNAP}, and combined with 300 local supernovae uniformly
distributed in the $z=0.03-0.08$ range. We study how the errors of the
parameter of most interest, $w_{\rm pivot}$, change as the individual SNa errors vary. We
again assume that the statistical error per SNa scales with redshift as

\begin{equation}
\sigma_m(z)=\sigma_m(0) + A z,
\label{eq:sigma_of_z}
\end{equation}

\noindent where $\sigma_m(0)=0.10$, $0.15$ or $0.20$, and we now let $A$ vary
in order to ascertain sensitivity to this parameter

As can be seen in
Fig.~\ref{fig:sig_wpivot_vs_A}, the constraints on $w_{\rm pivot}$ are
dependent on both $\sigma_m(0)$ and $A$; we can fit the constraint via an
approximate relation

\begin{equation}
\sigma(w_{\rm pivot})\approx 0.17\,(2\sigma_m(0) + A).
\end{equation}

This equation shows that the slope of the statistical error vs.\ redshift
relation, $A$, contributes to $\sigma(w_{\rm pivot})$ one half as much as the
intercept of the same relation, $\sigma_m(0)$. For  example, a 20\% increase
in the SNa error between the redshift of 0 and 1 (so that $A=0.2\sigma_m(0)$)
will lead to a 10\% increase in error associated with the pivot value of the equation of state. 

\section{Conclusions}\label{sec:conclusions}

 SNe Ia are currently the strongest cosmological probes of dark energy, and are
likely to remain the most solid source of information in the near future. This
implies that control of the statistical errors is crucial.  In particular, if
we are ever to address the central question of cosmology, 'Is the dark energy
due to a cosmological constant?', we need to be able to unambiguously determine
that $w \ne -1$ at some, or all redshifts.  The extent to which SNe Ia
measurements will allow such a determination has been our prime concern in this
analysis

Figures 4 and 5 represent our primary results in this regard.  
Figure 4 makes clear that the model-independent constraints on $w(z)$, for a
two-parameter class of deviations around $w(z)=-1$ (the two parameters are
$w_0$ and $w_a$; see Eq.~(\ref{eq:wz_Linder})), are quite sensitive to
the  measurement uncertainty in supernova magnitudes.  Moreover, both Figure 4 and
Fig.~\ref{fig:sig_wpivot_vs_A} demonstrate that it is particularly important to
attempt to maintain control over measurement uncertainties for higher redshift
supernovae.   As Figure 4 also demonstrates, allowing for possible variations in $w$ with redshift  implies to reduce the $ 95 \%$ confidence limit uncertainty in $w(z)$ for a fit near $w_0 =-1$ and
$w_a=0$ significantly below 10 $\%$ in $w_0$ will be challenging.

This latter point is even more important when considering whether one might
utilize the measured value of $w(z)$ at the pivot point to attempt to discern
some time at which $w \ne -1$.  Redshift-dependent uncertainties in supernova
magnitudes can easily almost double the inferred uncertainty in $w$ at this
point.  Moreover, only if the planned large scale SNe surveys can maintain a
uniform magnitude uncertainty per supernova, $\sigma_m$, less than $0.10$ can
we hope to derive a $95 \%$ confidence limit uncertainty in $w_{\rm pivot}$ of
less than about $ 10 \%$ (see Fig.~\ref{fig:sig_wpivot_vs_A}) which itself may
not be sufficient to distinguish some non-standard dark energy models from a
cosmological constant.

It is not clear whether resolving the nature of dark energy is a 10-year or a
100-year problem. The answer partly depends on how much information
measurements of dark energy can provide. Here we have addressed the former
problem by studying the requirements of magnitude errors in a SNa Ia survey.
Our results may be viewed as discouraging, or they may be viewed as inspiring
for those observers who enjoy a demanding challenge.  
It is already becoming clear that, in order to have hope of detecting a deviation
from the cosmological constant scenario,  we need nature 
to be kind by producing a non-negligible deviation in $w(z)$ at some redshift, {\it and} we
need to control SNa magnitude errors, and systematics in particular, to high precision. 
If either one of these requirements is not met, may have to rely on theorists to
understand the nature of dark energy, which is option over which we may have much less
control.





\end{document}